\documentstyle[amssymb,prb,aps,epsfig]{revtex}

\begin{document}
\twocolumn[\hsize\textwidth\columnwidth\hsize\csname
@twocolumnfalse\endcsname

\title{Charge ordering and phase separation \\
       in the infinite dimensional extended Hubbard model}
\author{Ning-Hua Tong$^{1}$, Shun-Qing Shen$^{2}$,Ralf Bulla$^{1}$}
\address{
$^{1}$ Theoretical Physics III, Center for Electronic Correlations and Magnetism,\\
Institute of Physics, University of Augsburg, D-86135 Augsburg, Germany \\
$^{2}$Department of Physics, The University of Hong \ Kong, Pokfulam, HongKong, China\\
}
\date{
\today%
%
}

\maketitle

\begin{abstract}

We study the extended Hubbard model with both on-site ($U$) and nearest neighbor ($V$) Coulomb repulsion using
the exact diagonalization method within the dynamical mean field theory.
For a fixed $U$ ($U=2.0$), the $T-n$ phase-diagrams
are obtained for $V=1.4$ and $V=1.2$, at which the ground state of $n=1/2$ system is charge-ordered and
charge-disordered, respectively.
In both cases, robust charge order is found at finite temperature and in an extended
filling regime around $n=1/2$. The order parameter changes non-monotonously with temperature.
For $V=1.4$, phase separation between charge-ordered and charge-disordered phases is observed in the low
temperature and $n < 0.5$ regime. It is described by an "S"-shaped structure of the
$n-\mu$ curve. For $V=1.2$,
the ground state is charge-disordered, and a reentrant charge-ordering transition is observed for
$0.42 < n < 0.68$. Relevance of our results to experiments for doped manganites is discussed.

\end{abstract}

\vskip2pc]

\section{Introduction}

Charge-ordering is a fascinating research topic in condensed
matter physics. In recent years, charge order and the related spin
and orbital order in doped manganites have attracted much interest.
In the system $La_{1-x}Ca_{x}MnO_{3}$ ($x>0.5$), charge order in the
ground state is enhanced as $x$ increases. The corresponding transition
temperature $T_{\rm co}$ is higher than the N\'{e}el temperature \cite{Cheong1},
and there are strong charge-order fluctuations at temperatures above $T_{\rm co}$
(Ref. \cite{Kim1}). In particular, a reentrant charge-ordering transition has been
observed in systems such as $Pr_{0.65}(Ca_{0.7}Sr_{0.3})_{0.35}MnO_{3}$
(Ref. \cite{Tomioka1}) and $LaSr_{2}Mn_{2}O_{7}$ (Ref.\cite{Kimura1,Li1}).
The origin of these charge-ordering transitions lies in the complex interplay between
orbital and lattice degrees of freedom in the manganites, and is presently under
intensive study. In other systems such as the heavy fermion system $Yb_{4}As_{3}$
(Ref. \cite{Ochiai1}), the quasi-one dimensional material $NaV_{2}O_{5}$
(Ref.\cite{Vojta1,Hubsch1}), and the superconducting layered organic
molecular crystal $\kappa -(BEDT-TTF)_{2}X$ (Ref.\cite{Mori1}), charge-ordering
is closely related to the specific properties of the system.

The charge ordering in the above stated systems has different
physical origins, and cannot be explained within a single theory.
From a theoretical point of view, an obvious cause of
charge-ordering is the
short-range Coulomb repulsion between electrons. The simplest model that
includes this interaction is the extended Hubbard model which contains
the kinetic term and the on-site and nearest-neighbor  Coulomb repulsion.
Despite the simplicity of this model, recent studies found that it can explain some
characteristics of the experimental observations in doped manganites\cite{Merino1,Pietig1,Kagan1}.
 These studies revealed that many interesting
effects arise simply from pure Coulomb repulsion and charge fluctuations. In
particular, using dynamical mean-field theory (DMFT), Pietig $et$ $al.$ found that
the quarter-filled extended Hubbard model exhibits a reentrant charge-ordering transition
near a critical value $V_{c}$ of the nearest-neighbor repulsion\cite{Pietig1} (Fig. 1). It coincides
with what was observed in the doped manganites $%
Pr_{0.65}(Ca_{0.7}Sr_{0.3})_{0.35}MnO_{3}$ (Ref. \cite{Tomioka1}) and $%
LaSr_{2}Mn_{2}O_{7}$ (Ref.\cite{Kimura1,Li1}). This indicates that some
properties of charge order may be independent of the concrete
microscopic mechanism, and hence
can be studied using simplified models such as the quarter-filled extended Hubbard
model. Along this line, other authors also studied this problem \cite{Tuan1,Hellberg1,Calandra1,Hoang1}.
In lower dimensions, the model has also been extensively studied in various contexts
using different methods\cite{Sengupta1}.

Experimentally, in the doped manganites, charge-ordering was observed not
only at $x=0.5$, but also in a broad doping regime. Also the reentrant behavior was
observed in systems away from quarter-filling. In fact, $%
Pr_{0.65}(Ca_{0.7}Sr_{0.3})_{0.35}MnO_{3}$ has an electron filling of $%
n=0.65>0.5$. Another feature of the experimental observation is the inhomogeneous
 coexistence of charge-ordered and charge-disordered phases. In the system $%
R_{1-x}Ca_{x}MnO_{3}$ $(R=La,$ $Nd,$ $Bi$ $etc.)$, many experiments showed
that phase separation (PS) between a ferromagnetic charge-disordered phase and
an antiferromagnetic charge-ordered phase exists at dopings ranging from $%
x=0.33 $ (Ref.\cite{Ibarra1}) to $x=0.82$ (Ref.\cite{Liu1}). Although detailed
explanations should take into account the spin and orbital degrees of freedom,
simplified models also give remarkably similar results. In the spinless fermion
 model, when electrons (holes) are added to a half-filled system ($n=1/2$),
 an electron (hole) rich charge-disordered phase is  separated from the
 charge-ordered background\cite{Uhrig1}.

In this paper, motivated by the experimental observations as well as by the
work of Pietig $et$ $al.$\cite{Pietig1}, we study the properties of a single-band extended
Hubbard model away from but near quarter-filling. We confine our study to the
two-sublattice case and do not consider a possible incommensurate ordering.
We are interested in the following two issues of charge-ordering. First, the
stability of the reentrant charge-order transition in the
whole doping regime. Second, the possibility of PS between
charge-disordered and charge-ordered phases. These issues are relevant to the charge-ordering
and PS phenomena observed in doped
manganites. Theoretical studies in the past focused only on
fillings at or near $n=1.0$ and $n=0.5$. Despite the intensive studies, these
two issues remain unclear.

\begin{figure}[ht]
\vspace{-1.0cm}
\begin{center}
\psfig{figure=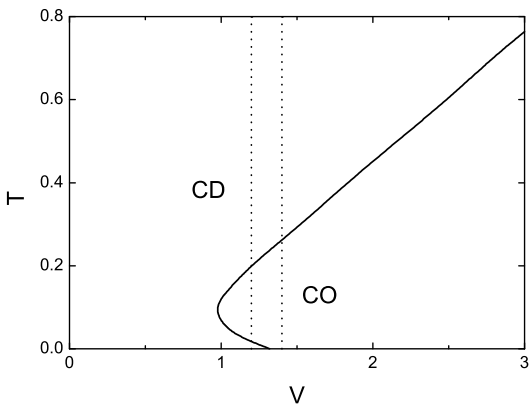,width=0.48\textwidth}
\vspace{-0.8cm}
\caption{Schematic $T-V$ phase diagram for the extended Hubbard model at $U=2.0$ and $n=1/2$ (taken
from Ref.\cite{Pietig1}, with $V$ rescaled.). CO and CD denote charge-ordered and charge-disordered,
respectively. The vertical dashed lines mark out the positions of $T-n$
planes studied in this paper, i.e., $V=1.2$ and $V=1.4$, respectively.}
\label{fig1}
\end{center}
\end{figure}

Using DMFT together with the exact
diagonalization technique, we study the paramagnetic phase diagram in two $T-n$ planes
for a fixed $U$-value $U=2$. As shown in Fig. 1, these two planes cross the
$T-V$ plane at $V=1.2$ and $V=1.4$, which are larger and smaller than $V_{c}$, respectively ($%
V_{c}$ is the critical value of the nearest-neighbor repulsion for a quarter-filled system,
which separates charge-ordered and charge-disordered ground states. $
V_{c}\approx 1.32W$ (Ref\cite{Pietig1}), see Fig. 1.). We found that a
reentrant charge-ordering transition exists in an extended regime
 of the electron
density $n$ near quarter-filling.

For $V>V_{c}$, the order-paramter of
charge-ordering changes non-monotonously as
temperature decreases. The ground state is still
charge-ordered for $n>0.5$ regime and charge-disordered for $n<0.5$. In the former regime,
We also find the PS between charge-ordered and charge-disordered phases.

For $V$ a little smaller than $%
V_{c}$, the ground state is charge-disordered for any $n$. The charge ordering exists
only at finite temperatures and is most robust in the $n>0.5$ regime.
At the lower
critical temperature of the
reentrant transition, the order-parameter disappears
rather abruptly. Phase diagrams in these two planes are plotted.

In Sec. II, we describe the model and the method used in this work.
In Sec. III, our results for $V=1.4W>V_{c}$ are presented.
Phase separation
between charge-ordered and charge-disordered phases are discussed. Their
thermodynamical structure is compared with other kinds of PS's.
Our results for $V=1.2W<V_{c}$ are presented in Sec. IV.
A summary is given in Sec. V.

\section{Model and method}

The Hamiltonian of the single-band extended Hubbard model has the form

\begin{eqnarray}
H & = & -t\sum\limits_{\left\langle i,j\right\rangle \sigma }c_{i\sigma }^{\dagger}c_{j\sigma }
       +U\sum\limits_{i}n_{i\uparrow }n_{i\downarrow}   \nonumber\\
  & + & V\sum\limits_{\left( i,j\right) }n_{i}n_{j}-\mu \sum_{i}n_{i}.  \label{1}
\end{eqnarray}
In Eq.(1), $\sum_{\left\langle i,j\right\rangle }$ indicates
the sum over nearest-neighbor sites $i$ and $j$ independently. $\sum_{\left( i,j\right) }$
indicates the sum over nearest pairs. Hence there is the relation $\sum_{\left(
i,j\right) }=1/2\sum_{\left\langle i,j\right\rangle }$. $U$ and $V$ are
on-site and nearest-neighbor Coulomb repulsion, respectively, and
 $\mu $ is the
chemical potential. Here we use a Bethe lattice which produces a
semicircular density of
states for free electrons in the limit of large coordination
number. In this limit,
 the inter-site coupling terms in Eq.(1)
are replaced by the corresponding Hartree term.
 After a proper rescaling of $V$: $V\to V/Z$ (Here $Z$ is the number of nearest neighbors)
and neglecting the constant term, we obtain the following mean-field
Hamiltonian:

\begin{eqnarray}
H_{\rm mf} & = &
      -t\sum\limits_{\left\langle i,j\right\rangle \sigma }c_{i\sigma}^{\dagger}c_{j\sigma}
      +U\sum\limits_{i}n_{i\uparrow}n_{i\downarrow} \nonumber\\
           & - & \sum\limits_{i}(\mu-V\langle n_{i+\delta} \rangle)n_{i},
\label{2}
\end{eqnarray}
where $i+\delta$ is the nearest site of $i$. Note that our scaling is different from that
of Ref.\cite{Pietig1}, where the scaling $V\to 2V/Z$ is used. So in this paper the value of $V$
is twice as large as the corresponding one in Ref.\cite{Pietig1}.

To describe the charge-ordered phase, we divide the Bethe lattice
into two sublattices. Correspondingly, within DMFT, the
model Eq.(2) is
 mapped onto two uncoupled effective Anderson impurity models,

\begin{eqnarray}
H_{\Lambda, \rm imp}
 & = & \sum_{k=1,\sigma}^{N_S-1}{\left[ \epsilon_{\Lambda k}a^{\dagger}_{\Lambda k \sigma}
      a_{\Lambda k \sigma}+V_{\Lambda k} (a^\dagger_{\Lambda k \sigma}c_{\Lambda \sigma}
      + h.c. ) \right]}   \nonumber\\
 & + & U n^c_{\uparrow} n^c_{\downarrow}
-\left( \mu -V \left\langle n^c_{\overline{\Lambda}} \right\rangle \right) n^c_{\Lambda}.
\label{3}
\end{eqnarray}

Here, $\Lambda = A,B$ refers to the two sublattices
($\overline{A}=B$, $\overline{B}=A$) and
 $\left\{ \epsilon_{\Lambda k}, V_{\Lambda k} \right\}$
are effective parameters describing the bath. They are related to the Weiss function
${\mathcal G}_{\Lambda 0} ^{-1} \left( i \omega_{n} \right)$ through

\begin{equation}
{\mathcal G} ^{-1}_{\Lambda 0} \left( i \omega_{n} \right)_{map} =i\omega_{n} + \left( \mu-
V \left\langle n_{\overline{\Lambda}} \right\rangle \right) -
\sum_{k=1}^{N_{\rm S}-1}
\frac{V^2_{\Lambda k}}{i \omega_{n} - \epsilon_{\Lambda k}}.
\label {4}
\end{equation}
We use the full exact diagonalization method to calculate the impurity Green's function for this model. The
number of sites $N_{\rm S}=5$ and $N_{\rm S}=6$ are found to be
sufficient for the calculations in this paper. The free
density of states is given by:

\begin{equation}
D(\epsilon)=\frac{2}{\pi W^2} \sqrt{W^2-\epsilon^2} \qquad (|\epsilon|<W) \, .  \label{5}
\end{equation}
We set $W=1$ as the energy unit.
The DMFT self-consistency equations for the Bethe lattice
are the given by

\begin{equation}
{\mathcal G}^{-1}_{\Lambda 0}(i \omega_{n})_{dys}=i\omega_{n}+\mu-V \left\langle n_{\overline{\Lambda}}
\right\rangle -\frac{W^2}{4} G_{\overline{\Lambda}}(i \omega_{n}) ,
\label{6}
\end{equation}
where $\Lambda=A$, $B$. Equations (3), (4), and (6) form
a set of closed self-consistent DMFT equations. After
the Green's function and Weiss function are calculated, the new set of effective parameters for the impurity
model is obtained via a minimization procedure\cite{Caffarel1}:

\begin{equation}
d  =  \frac{1}{n_{max}+1} \sum_{n=0, \Lambda}^{n_{max}}
      \left| {\mathcal G}^{-1}_{\Lambda 0}(i\omega_{n})_{map} - {\mathcal G}^{-1}_{\Lambda 0}(i\omega_{n})_{dys}
 \right| ^{2}.
\label{7}
\end{equation}

The self-consistent equations are solved iteratively.
After the iteration converges,
we calculate the electron densities on the two sublattices, $n_{A}$ and $n_{B}$, which in turn
give the average electron density
$n=\left( n_{A} +  n_{B}\right)/2$ and the order parameter $\left| n_{A}-n_{B} \right|$.
The contributions to the total energy per lattice site
$E=E_{T}+E_{U}+E_{V}$ from different parts of the Hamiltonian are also calculated.
The kinetic energy $E_{T}$, on-site repulsion energy
$E_{U}$, and the inter-site repulsion energy $E_{V}$ are given by:

\begin{equation}
E_{T}  =   \frac{1}{\beta} \sum_{n, \Lambda} \xi_{ \Lambda} \left( i \omega_{n} \right) 
          G_{\Lambda}\left( i\omega_{n} \right) e^{i\omega_{n} 0^{+}},
\label{8}
\end{equation}
where

\begin{equation}
\xi_{\Lambda} \left( i \omega_{n} \right)  =
                 i\omega_{n} +\mu -{\mathcal G}^{-1}_{\Lambda 0}\left( i \omega_{n} \right) \;
                 \left( \Lambda = A , B \right) , \label{9}
\end{equation}

\begin{eqnarray}
E_{U} & = & \frac{1}{2 \beta} \sum_{n, \Lambda}  \nonumber\\
      && \left[ {\mathcal G}^{-1}_{\Lambda 0} \left( i\omega_{n} \right)
         -G^{-1}_{\Lambda} \left( i\omega_{n} \right) \right]
         G_{\Lambda} \left( i\omega_{n} \right) e^{i \omega{n} 0^{+}}, \label{10}
\end{eqnarray}

and
\begin{equation}
E_{V}=\frac{V}{2} n_{A} n_{B} \, . \label{11}
\end{equation}

Phase separation has been studied extensively for
models of  strongly correlated electrons, such as the
Hubbard model\cite{Michielsen1,Dongen1,Zitzler1}, the $t-J$ model\cite{Emery1,White1}, the Falicov-Kimball model
\cite{Freericks1}and the double-exchange model\cite{Shen1,Yunoki1,Moreo1}. A standard
criterion for the PS is the discontinuous jump in the curve of electron density $n$ versus chemical
potential $\mu$ (Ref.\cite{Moreo2}). However, in the DMFT study of double-exchange systems, we have
observed PS also
through the multiple-valued structure in a continuous $n-\mu$ curve\cite{Tong1}.
As described in the next section,
we find  in a certain temperature regime for $V=1.4$ an "S" or
"Z"-shaped multiple-valued structure in the curves of $n_{A}$, $n_{B}$ and $n$ versus $\mu$. Such curves
contain the full information about the PS, including the metastable phase and the first-order
phase transition line. To do this, we introduce\cite{Tong1} a self-consistently
determined quantity $\mu '$,

\begin{equation}
\mu '=\mu-\lambda \left( n_{A} -A \right), \label{12}
\end{equation}
where $\lambda$ and $A$ are tunable parameters. The value of $n_{A}$ is dependent on
the chemical potential $\mu$ through $n_{A}=F( \mu )$ and the functional dependence $F(\mu )$
is  determined by the DMFT calculations. For a given $\mu$ and
in each DMFT iteration, we first calculate $\mu '$ and use $\mu '$ instead
 of $\mu$ in the ordinary
DMFT scheme to produce the local Green's function, and then
extract the $\epsilon_{\Lambda k}$ and $V_{\Lambda k}$.
After that, the new value of $n_{A}$ is calculated. The iteration is carried on until convergence is reached.
This is equivalent to simultaneously solve the DMFT equation and the following equation,

\begin{equation}
n_{A} = F {\left[ \mu -\lambda \left( n_{A} -A \right) \right]}. \label{13}
\end{equation}

If the function $F( \mu )$ has a multiple-valued regime, by selecting appropriate parameters $\lambda$ and
$A$, the $n_{A}- \mu $ curve self-consistently determined by Eq. (13) may become
single-valued. In this way, our calculation avoids the numerical instabilities induced by
the multiple-valued structure of $n- \mu $ curve. For each $\mu$, we first solve Eq. (13)
together with the DMFT equations, then calculate the thermodynamic quantities $Q$ that we are
interested in. After the data is obtained for each $\mu$, we plot the quantities $Q$ with respect to
the argument $\mu '$ to recover the physical curves $Q=F_{Q} \left( \mu ' \right)$ that
correspond to the orginal Hamiltonian. The final results should be independent of the parameters
$\lambda$ and $A$, if only they are in an appropriate regime. In the framework of DMFT,
this transformation scheme has been used in the study of  PS in the
double exchange model\cite{Tong1}
and of the Mott-Hubbard transition in the Hubbard model\cite{Tong2}.

\section{Results and discussion}

\subsection{Phase separation: $V=1.4 > V_{c}$}

In this section, we discuss our results for $V=1.4$, which is a little larger than the
zero-temperature critical value $V_{c}\approx 1.32$ (See Fig.1). For this interaction strength,
as is shown in Fig. 1, the exactly quarter-filled system has a charge-ordered ground state. The charge
order persists up to $T \approx 0.26$, and no reentrant transition was found. In order to study
the system away from quarter filling, at a lower temperature $T=0.05$,
 we change the chemical potential and calculate charge densities.
 Fig.2(a) shows our result for the sublattice charge densities
 $n_{A}$, $n_{B}$, and average
charge density $n$ as functions of $\mu$. As we have expected, charge order exists in some
finite regime of doping around $n=1/2$: $0.44<n<0.73$. This regime is not symmetric about $n=1/2$,
since the model (1) does not have particle-hole symmetry at this point.
 This differs
from the spinless fermion model\cite{Uhrig1}. Here,
the charge-ordered regime extends  to larger densities,
where the inter-site Coulomb repulsion is more effective. However, the value of
$\left| n_{A}-n_{B} \right|$ is largest at $n=1/2$.

\begin{figure}
\vspace{-1.0cm}
\begin{center}
\psfig{figure=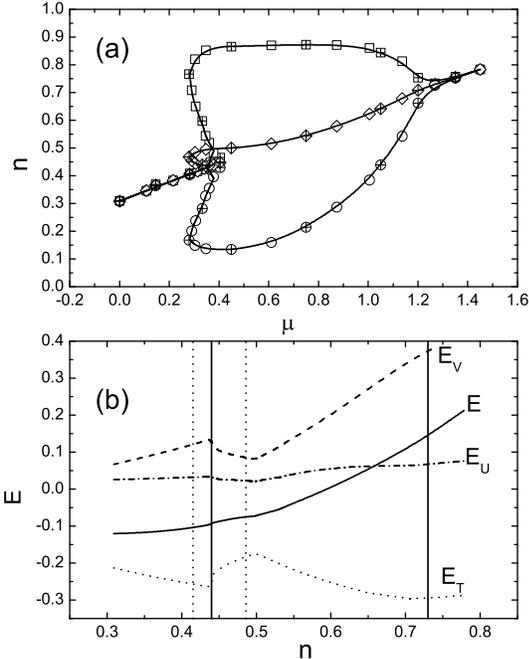,width=0.48\textwidth}
\vspace{-0.8cm}
\caption{(a) electron density $n$ versus chemical
potential $\mu$ for U=2.0, V=1.4 and T=0.05.
The squares, dots, and diamonds are for $n_{A}$,
$n_{B}$, and $n$, respectively. Calculation
results with Ns=6 (hollow symbols) and Ns=5
(cross-filled symbols) agree well.
Lines are for guiding eyes.
(b) Contributions to the energy per lattice
site as functions of $n$: $E_{T}$, $E_{V}$, $E_{U}$ and $E$ denote kinetic
energy, nerest neighbor repulsion energy, on-site repulsion energy, and the total energy,
respectively. The regime between the dotted vertical lines is unstable
towards phase separation. Charge order exists between the solid
vertical lines.}
\label{fig2}
\end{center}
\end{figure}

One dominant feature of Fig. 2(a) is the "S"-shaped multiple-valued structure in the $n-\mu$,
$n_{A} -\mu$ and $n_{B} -\mu$ curves. This is direct evidence for PS in the extended
Hubbard model near quarter filling. When $n$ increases, the system first
goes from a charge-disordered phase into a
charge-ordered phase through a first-order transition, which
occurs in the regime $n=0.4\sim 0.5$. As $n$ increases further, the charge order disappears
continuously at $n\approx 0.73$. The results obtained using $N_s=5$ and $N_s=6$ agree very well in most
part of the curve. There is only  a slight deviation near the second-order transition point at $n=0.44$.
This indicates that for this problem, $N_s=5$ is sufficient to obtain
 qualitatively
correct conclusions. In the following, all our results are from $N_s=5$ calculations.

In the multiple-valued regime of the $n-\mu$ curve, the DMFT
self-consistency equations have three solutions
for a fixed $\mu$. We use the thermodynamical grand potential $\omega \left( T, \mu \right)$ to
evaluate the relative stability of these solutions:

\begin{equation}
\omega \left( T, \mu \right) = \omega \left( T, \mu_{0} \right)-\int_{\mu_{0}}^{\mu} n
\left( T, \mu ^\prime \right) \, d \mu ^\prime. \label{14}
\end{equation}
The actual first-order transition point $\mu_{c}$ is determined by a Maxwell construction,
i.e., by solving the equation
$\omega_{1}\left( T, \mu_{c} \right)=\omega_{2}\left( T, \mu_{c} \right)$. At $\mu=\mu_{c}$, two
phases coexist, with their respective volumes determined by the nominal electron density of the system.
At $T=0.05$, one of the coexisting phases is charge-disordered with
$n_{1}=0.415$, the other one is charge-ordered with $n_{2}=0.486$. The third solution with
intermediate $n$ is charge-ordered, but due to its negative compressibility and highest grand
potential, it is unstable with respect to the others. For $\mu$ away from but near
$\mu_{c}$, there is only one stable phase in the system, either ordered or disordered.
The other two solutions have higher grand potential and are metastable.
These metastable phases may be detected by hysteresis experiments.

$n=0.44$ is the lower critical density at which charge ordering occurs. At this point, the $n-\mu$ curve
turns backwards sharply, and  $n=-\partial\omega \left( T, \mu \right) /\partial\mu$
is continuous while $\partial n/ \partial\mu$ is discontinuous.
Therefore it is identified as a second-order transition point between charge-disordered and
charge-ordered phases. In this respect, it is same as the $n=0.73$ point. However, due to the higher
grand potential, the second-order charge ordering transition only exists at metastable level.
To understand the PS better, we show the $n$-dependence of the energy in Fig. 2(b).
The on-site repulsion energy $E_{U}$ has a small contribution. It does not
 change much at the boundary to the charge ordering regime. This is an indication that for $U=2$, double
occupancy is small in the doping range around $n \sim 0.5$, and that it plays a minor role for PS.
In contrast, the kinetic energy $E_{T}$ and inter-site repulsion energy $E_{V}$ are
more sensitive to the charge density and long-range charge ordering. If there is no charge ordering,
$E_{V}$ behaves as $E_{V} \propto n^2$, while $E_{T}$ recovers the behavior of the
Hubbard model; this means that  with increasing $n$, it decreases in the low filling area and increases near
half filling.

In Fig. 2(b), it is seen that the charge order in the regime $0.44<n<0.73$ strongly
reduces $E_{V}$ while it increases $E_{T}$ with respect to their value in disordered phase. The charge ordering
transition at $n=0.44$ causes drastic changes in $E_{V}$ and $E_{T}$, which are then naturally
related to the occurence of PS. In contrast, at the other transition point $n=0.73$,
the energies change more smoothly. Because the opposite contributions from $E_{V}$
and $E_{T}$ almost cancel, the total energy $E$ increases monotonously
and smoothly with increasing $n$. However, the quick change of $E_{V}$ and $E_{T}$ near $n=0.44$
and their competition leads to a small convex part in the $E-n$ curve, where $\partial ^2E/\partial n^2 <0 $.
When taking into account the entropy contribution to the free energy, the convex structure will be
enforced by a small amount $TS \sim 10^{-2}$. Therefore, at the temperature $T=0.05$, the $E-T$ curve
already represents the behavior of free energy $F \left( T, n \right)$. The convex structure observed
here is therefore consistent with PS obtained using the $n- \mu$ criterion. From Fig. 2(b),
it is clear that PS is closely related to the charge ordering transition. It
directly results from the charge-ordering-induced strong competition between $E_{V}$ and $E_{T}$.

\begin{figure}
\vspace{-1.0cm}
\begin{center}
\psfig{figure=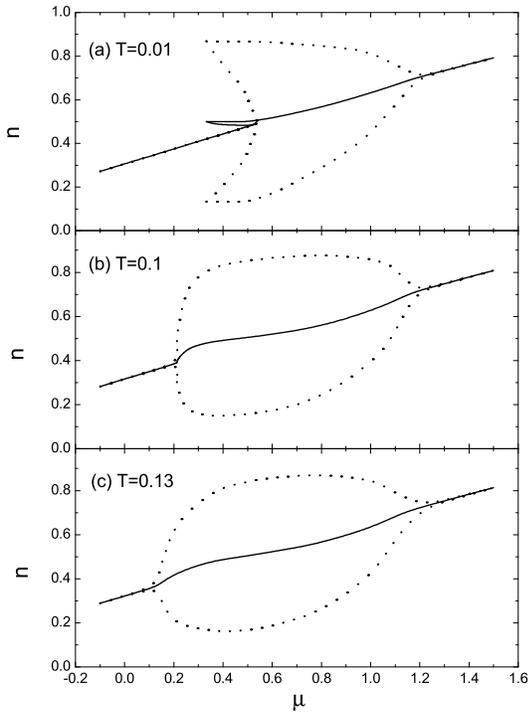,width=0.48\textwidth}
\vspace{-0.8cm}
\caption{Electron density $n$ versus chemical
potential $\mu$ for $U=2.0$, $V=1.4$, and (a)
$T=0.01$, (b) $T=0.1$, (c) $T=0.13$. The upper and
lower dotted lines and the solid line are for
$n_{A}$, $n_{B}$, and $n$, respectively.
}
\label{fig3}
\end{center}
\end{figure}

In the following, we study the temperature dependence of charge order and PS.
In Fig. 3, three $n-\mu$ (as well as $n_{A} - \mu$ and $n_{B} - \mu$) isotherms are shown
for $T=0.01$, $0.1$, and $0.13$. Up to temperature $T=0.13$, the charge order is quite robust
in the intermediate density regime from $n \sim 0.3$ to $n \sim 0.7$. In contrast, PS
is stable only at much lower temperatures. Compared with the $T=0.05$ curve in Fig.2(a),
the multiple-valued structure is more pronounced at $T=0.01$,
while it disappears at higher temperature.
As temperature decreases, the multiple-valued structure in $n-\mu$ curve is compressed along the
$n$ axis, but that in the $n_{A} -\mu$ and $n_{B} -\mu$ curves does not change much. As a result,
the average electron densities of the coexisting two phases, which are determined through Maxwell
construction, get closer and both move towards $n=0.5$.
On the other hand, as temperature increases,
the multiple-valued part of the $n-\mu$ curve shrinks along the
$\mu$ axis until it disappears
at $T\approx 0.1 $(Fig.3(b)). At this temperature, PS disappears and the slope
of the $n-\mu$ curve diverges at the second-order charge ordering transition, leading to strong
fluctuations in charge density as well as in the order parameter $\left| n_{A} - n_{B} \right|$.
The critical end point of the PS is estimated to be $n \approx 0.39, \, T \approx 0.1$.

It is interesting to note that the low temperature behavior of PS is not unique to this model.
Similar behavior has been observed in the DMFT study of first-order phase transitions in other
strongly correlated electron models. The ferromagnetic-paramagnetic (FM-PM) PS in
the double-exchange model\cite{Tong1} and the Mott-Hubbard metal-insulator transition in the half-filled
Hubbard model\cite{Tong2} are both typical first-order phase transitions. Within DMFT, they are described
by continuous "S" or "Z"-shaped curves that are very similar with the $n-\mu$ curves presented here.
As the temperature decreases, all these structures are compressed in the vertical direction.
The "order parameter" of the first-order transitions ($\left| n_{FM} -n_{PM} \right|$ for FM-PM
PS, and double occupation difference $\left| D_{Met} -D_{Ins} \right|$ for the
Mott-Hubbard transition) obtained from a Maxwell construction reduces to zero in the limit
of $T \rightarrow 0$.

Here, though the lowest temperature that we study is $T=0.01$, the main tendency
is clearly that as temperature decreases, the upper two branches of the "S"-shaped curve (as shown in
Fig. 3(a)) tend to merge. Hence we expect that as $T \rightarrow 0$, the density
difference between the two coexisting phases  reduces to zero, similar with our previous
findings in other systems. The two coexisting phases have the same density $n=0.5$ and the same energy.
In such a scenario, the ground state of the $V=1.4$ system is singular at $n=0.5$.
At this point, a charge-disordered phase and a charge-ordered phase, both with avarage density $n=0.5$,
coexist in random volume proportion. An infinitesimal amount of
additional holes in the system will destroy
the phase-separated ground state and turn it into charge-disordered state, while
electrons will turn it into charge-ordered state with $\left|n_{A} -n_{B} \right| \approx 0.7$. Again,
this is different from the spinless fermion system, for which both additional holes and electrons doped
 into the charge-ordered ground state at $n=0.5$ cause PS\cite{Uhrig1}.

In Fig. 4, the $T-n$ phase diagram for $U=2.0$ and $V=1.4$ is shown. It is seen that charge order
is rather robust in the filling regime around $n=0.5$. At zero temperature, the charge-ordered ground
state extends from $n=0.5$ to $n \approx 0.69$. This range first expands, and then shrinks as
temperature increases, leading to a reentrant transition of charge ordering in certain doping regimes.
 The highest charge ordering transition temperature $T_{\rm CO,max} \approx 0.28$ is reached at $n=0.5$.
The PS area lies near the lower transition temperature line on the left side.
When both  the nominal electron filling and the temperature
 lie in this area, a charge-disordered phase with
lower electron density will be separated from a charge-ordered phase with higher electron density.
The two boundaries of this coexisting area, as shown by the solid lines in the figure, meet at two
end points. The finite temperature end point is located at ($T_{c} \approx 0.1, n_{c} \approx 0.39$),
which is easily seen for ordinary first-order phase transitions such as liquid-gas transition. The other
end point is located at zero temperature ($T=0, n=0.5$), which is a common feature of the first-order
PS described by DMFT\cite{Tong1,Tong3}. The second-order
charge ordering transition line extends to zero temperature. Between this line (thin line in Fig. 4)
and the dashed line lies an  unstable area where the compressibility is negative. Metastable phases,
either charge-disordered or
charge ordered, exist in the two patches between the unstable area and the boundaries. They are
intrinsic features of first-order phase transitions and important for the properties of materials
near PS. In doped manganites, metastable phases have been observed through the hysteresis of magnetization
and resistivity with respect to temperature\cite{Schiffer1,Babushkina1}, as well as
through the resistivity relaxation phenomenon\cite{Anane1}. The study of metastable phases has
provided valuable information about the PS in doped manganites.

\begin{figure}
\vspace{-1.0cm}
\begin{center}
\psfig{figure=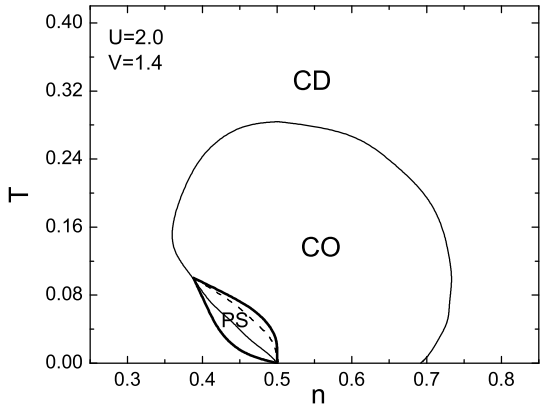,width=0.48\textwidth}
\vspace{-0.8cm}
\caption{Phase diagram in the $T-n$ plane
for $U=2.0$ and $V=1.4$. The thin solid line is the
second-order transition line. CO and CD denote charge-ordered and charge-disordered phase,
respectively. PS denotes phase
separation between charge-ordered and
charge-disordered phases. The thick solid lines
are the boundaries of the coexisting regime. In this regime,
between the dashed line and the thin line lies a charge-ordered phase
with negative compressibility.}
\label{fig4}
\end{center}
\end{figure}

Electronic PS has been studied extensively in various models such
as the Hubbard model, the $t-J$ model, and the double-exchange model, $etc$. Most of the PS
scenarios discussed so far
rely on the magnetic exchange mechanisms. Up to now, PS induced by pure Coulomb repulsion is observed
only in the spinless fermion model\cite{Uhrig1}.
Our results show another example of PS caused by Coulomb repulsion only.
In general, the Coulomb repulsion works against the PS, since a
phase-separated state has a higher potential energy than the charge uniform state. Here we
find that PS can also be driven by the charge-ordering transition
which is in turn induced by pure Coulomb repulsion.
When the charge orders to avoid the strong nearest-neighbor Coulomb repulsion in the high density regime,
the Coulomb potential is strongly reduced
which then allows for PS.
It should also be noted that the PS described here exists mainly at finite temperatures, and comes from
the effect of thermal fluctuations. In this respect, it is different from the ground-state PS of the
spinless fermion model.

In the following we discuss the effect of long-range (beyond the nearest neighbor) Coulomb repulsion on the
stability of PS between charge-disordered and charge-ordered phases.
It is generally believed that long-range Coulomb repulsion will
suppress a complete PS in the system, leading to an
inhomogenious distribution of the electron
density. Here we confine our discussion to the two-sublattice case and do not
consider any incommensurate ordering. Taking into account the long-range Coulomb repulsion, the total
inter-site part of the Hamiltonian (including the nearest-neighbor contribution) can be formulated as

\begin{equation}
H_{\rm int}=\sum_{i,j} V_{ij}n_{i}n_{j}, \label{15}
\end{equation}
In infinite dimensions, Eq. (15) reduces to its Hartree form.
After performing appropriate scalings for the parameters $V_{i,j}$,
the effective mean-field Hamiltonian has the form:

\begin{eqnarray}
H_{\rm int} & = & \sum_{i \in A} n_{iA} \left( V_{1}\left\langle n_{B} \right\rangle + V_{2}
                   \left\langle n_{A} \right\rangle \right)    \nonumber\\
            & + & \sum_{i \in B} n_{iB} \left( V_{1}\left\langle n_{A}
 \right\rangle + V_{2} \left\langle n_{B} \right\rangle \right). \label{16}
\end{eqnarray}
The effect of long range repulsion reduces to
only the nerest-neighbor type and the next-nerest-neighbor type, which are represented
by $V_{1}$ and $V_{2}$, respectively. If we denote $\left\langle n_{A} \right\rangle = n+ \delta$ and
$\left\langle n_{B} \right\rangle = n- \delta$, then Eq. (16)
further
simplifies to

\begin{eqnarray}
H_{\rm int} & = & \sum_{i} \left( V_{1} +V_{2} \right)n\,n_{i} + \delta \left( V_{2}-V_{1} \right)
                  \sum_{i \in A} n_{iA}     \nonumber\\
            & + & \delta \left( V_{1} - V_{2} \right) \sum_{i \in B} n_{iB}.  \label{17}
 \end{eqnarray}

This means that the long-range Coulomb repulsion introduces two effects.
One is the enhancement of the average repulsion as shown in the
first term. The other is the frustration effect caused by the next-nerest-neighbor type repulsion.
In Eq.(17) this effect is reflected by the terms proportional to $V_{2} - V_{1}$.
The first term can effectively suppress the coexisting regime, but cannot destroy PS at
sufficiently low temperature\cite{Tong1}. In the low temperature limit, the density difference of two
coexisting phases reduces to zero, and the average repulsion will lose its effect on the PS.
In contrast, the frustration effect induced by long-range Coulomb repulsion may destroy the PS
completely, since PS crucially depends on the charge-ordering transition.
In particular, if the frustration is
so large that the effective nearest-neighbor
$V$ is less than $V_{c}$, as shown in the next section, there
is no PS at all.
Therefore we conclude that the long-range Coulomb repulsion may destroy the PS through
its frustration effect.

In the regime $V > V_{\rm c}$,
 the charge order becomes robust
 with increasing $V$, as shown
in Fig. 1. Therefore, we expect that as $V$ increases, the charge-ordered area in the $T-n$ phase
diagram will expand in both temperature and filling regime.
However, the main shape of the phase boundary
will remain unchanged. In particular, due to the close relation between PS
and the charge ordering transition,
the PS regime may well expand while keeping its main structure.

\begin{figure}
\vspace{-1.0cm}
\begin{center}
\psfig{figure=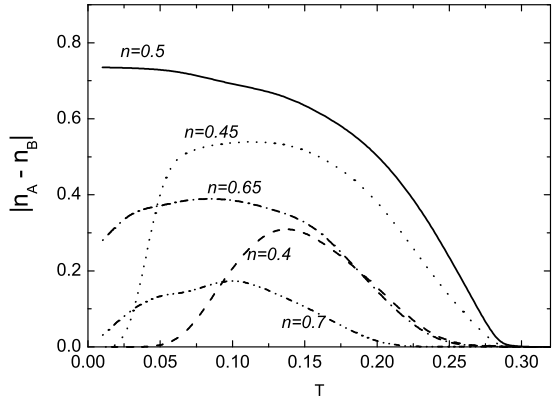,width=0.48\textwidth}
\vspace{-0.8cm}
\caption{The charge density
difference between sublattice A and B as functions of $T$
at $U=2.0$, $V=1.4$ for several electron fillings:
$n=0.4$ (dash), $0.45$ (dot),
$0.5$ (solid), $0.65$ (dash-dot), and $0.7$ (dash-dot-dot).}
\label{fig5}
\end{center}
\end{figure}

Besides PS, we are also interested in the properties of the charge order in this
 regime. In Fig. 5, the order parameter
$\left| n_{A} - n_{B} \right|$ is shown
as a function of temperature for different fillings. Except for
$n=0.5$,
$\left| n_{A} - n_{B} \right|$ shows a nonmonotonous behavior
 for all the fillings we studied.
It first increases and then decreases
upon lowering the temperature. In the filling regime $0.5 < n < 0.69$ where the ground state is charge
ordered, $\left| n_{A} - n_{B} \right|$ reduces to a finite value at $T=0$. For fillings outside
but close to this regime, the ground state is charge-disordered and
a reentrant transition occurs at finite temperature.
Compared with the reentrant transition,
the non-monotonous change of the order parameter is a more general phonomenon.
Experimentally, in systems with an ordered ground state but
 near the reentrant transition, such a non-monotonous change
of the order parameter may well be observed.

\subsection{Reentrant charge ordering: $V=1.2 < V_{\rm c}$}

In this section, we discuss the case of $V=1.2$, which is  smaller than
$V_{\rm c} \approx 1.32$. For this
repulsion strength, the ground state of the quarter-filled system is charge-disordered.
As temperature decreases, the system shows a disorder-order-disorder type reentrant transition\cite{Pietig1},
as shown in Fig.1. When the electron filling moves away from $n=0.5$, we find that
the ground state is still charge-disordered, and that a
 reentrant charge ordering transition exists
in an extended density regime $0.42 < n < 0.68$. In Fig.6, the curves for
 $\left| n_{A} - n_{B} \right|$
versus $T$ are shown for several fillings in this regime. It is seen that $\left| n_{A} - n_{B} \right|$
varies non-monotonously as temperature decreases, and drops to zero at some finite temperature.
Though the high temperature transition is continuous for all fillings, we find that the
reentrant transition at lower temperatures has a
different behavior for small and large fillings.
For the filling $n=0.4$ and $n=0.45$, $\left| n_{A} - n_{B} \right|$ changes smoothly at the reentrant
transition temperature, indicating that this transition is of second order. In contrast, in the curves for
$n=0.55$, $0.6$ and $0.65$, $\left| n_{A} - n_{B} \right|$ drops abruptly near the temperature $T=0.05$.
Accompanied with this rapid change of the order parameter
 an obvious slowing down of convergence is observed,
which reduces the numerical precision significantly.
A similar situation is found also for the filling $n=0.5$.
Since we have not found a hysteresis
 typical for first-order transitions, here we
would not conclude that it is a first-order phase transition. More detailed studies
are needed to elucidate this issue.

\begin{figure}
\vspace{-1.0cm}
\begin{center}
\psfig{figure=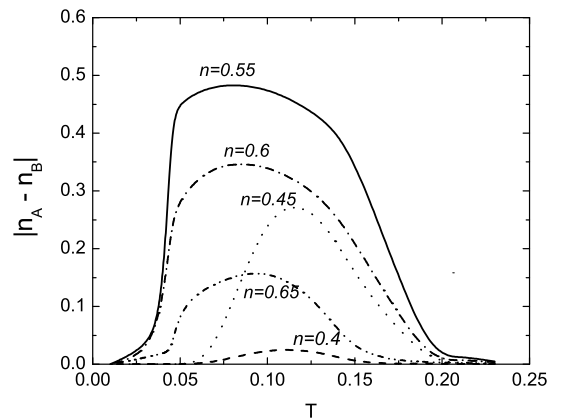,width=0.48\textwidth}
\vspace{-0.8cm}
\caption{The charge density
difference between sublattice A and B as functions of $T$ at
$U=2.0$, $V=1.2$ for several electron fillings:
$n=0.4$ (dash), $0.45$ (dot), $0.55$ (solid), $0.6$ (dash-dot),
and $0.65$ (dash-dot-dot).
}
\label{fig6}
\end{center}
\end{figure}

For both $V=1.4$ and $V=1.2$, the non-monotonous behavior of $\left| n_{A} - n_{B} \right|$ versus
$T$ originates from the large spin entropy of the paramagnetic charge-ordered phase. For the charge-ordered
phase, the two-fold spin degeneracy on the occupied sites contributes a total entropy
$\left( N/2 \right) \ln 2$, while for the charge-disordered state near $n=0.5$, the entropy increases from
zero with increasing $T$. So at low temperature, when the charge-ordered phase has a higher entropy than
the disordered phase, the charge order develops as temperature increases. At higher
temperature, where the entropy of the disordered phase exceeds that
of the charge-ordered phase, the charge order
is reduced with increasing temperature. This is the main reason that leads to the non-monotonous behavior
of the order parameter. Therefore, if the spin degeneracy of the charge-ordered phase is  destroyed
by the forth-order superexchange mechanism, it is doubted whether such a
reentrant behavior is still present.
Hellberg {\it et al.} \cite{Hellberg1}
carried out a finite-temperature Lanczos study on a
$4 \times 4$ lattice. They found, however,
that the reentrant transition is stable when the superexchange
effect is partly taken into account.

In Fig.7, the $T-n$ phase diagram for $U=2.0$ and $V=1.2$ is shown. The reentrant charge-ordering transition
exists in a regime from $n \approx 0.42$ to $n \approx 0.68$. Similar to the case of $V=1.4$, the charge ordering
regime is also asymmetric around $n=0.5$. Charge order
appears only at finite temperatures. Here, one important difference from the $V=1.4$ diagram is that
there is no PS near the reentrant transition line. For the fillings near
the two vertical boundaries of the charge-ordered area, the order parameter changes continuously to zero
at the lower transition temperature. This
continuous transition is indicated by a thick line Fig. 7. In the regime $0.5 < n < 0.65$,
the high temperature transition is continuous, while at low temperatures, the order parameter disappears more
abruptly. In the phase diagram we schematically denote such an abrupt transition by a dotted horizontal
line. Due to the severe critical slowing down of convergence at this transition, it is difficult to obtain
the transition temperature accurately.

Comparing Figs. 4 and 7, we see that the $T-n$ phase diagrams for $V = 1.4$ and $V=1.2$ are
topologically different. Near the critical $V_{\rm c}$, the $T-n$ phase diagram,
including the stability of PS and the reentrant transition, is very
sensitive to $V$. When $V$ increases from $V < V_{\rm c}$, both the $n$ and $T$ regime of the charge-ordered
area  expands, and the reentrant transition temperature decreases. But the ground state remains
disordered, as shown in Fig. 7. At $V=V_{\rm c}$, the ground state at $n=0.5$ is expected
to first turn charge-ordered.
When $V$ is even larger, the filling regime of the charge-ordered ground state extends towards larger
$n$, while keeping its left end-point $n=0.5$ unchanged. At the same time, a PS area where charge-disordered
and charge-ordered phases coexist emerges near the reentrant
transition line in the $n < 0.5$ regime.
In this way, the phase diagram shown in Fig. 7 for small $V$ evolves into that in Fig. 4 for large $V$.

Phenomena such as charge ordering, reentrant transition, and PS
have been observed experimentally in doped manganites.
The mechanisms for these phenomena are more complicated
and are topics of intensive research.
The interesting point here is that
 starting from an
electronic model that only takes into account the on-site and nearest-neighbor Coulomb repulsion,
we are able to obtain rich phase diagrams that include all these phenomena. Previous studies \cite{Pietig1}
indicated that the occurence of a reentrant charge ordering transition does not depend on details of the
localization mechanism. Here, when compared with other first-order phase transitions studied before, we
find that within DMFT, the structure of first-order transitions and their temperature
behavior are very similar. They are also independent of the specific mechanism. It would be interesting to
do such a comparison for those real materials where different first-order phase transitions occur.

\begin{figure}
\vspace{-1.0cm}
\begin{center}
\psfig{figure=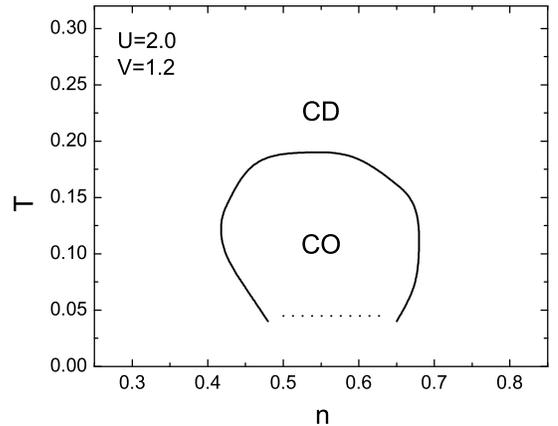,width=0.48\textwidth}
\vspace{-0.8cm}
\caption{Phase diagram in $T-n$ plane for $U=2.0$ and $V=1.2$.
The solid line is the second-order
transition line. CO and CD denote charge-ordered and charge-disordered phase,
respectively. The dotted line denotes an abrupt change of the order parameter.}
\label{fig7}
\end{center}
\end{figure}

\section{Summary}

In this paper, we studied the paramagnetic phase of the single band
extended Hubbard model near quarter filling. In the framework of DMFT, the effective Anderson impurity
 model is solved using the exact diagonalization technique. Based on the previous $T-V$ phase diagram for $n=1/2$
(Ref.\cite{Pietig1}), we concentrate on the phase diagram in $T-n$ planes for two specific value of the
inter-site repulsion
$V$: $V=1.4$ and $V=1.2$, located on the two sides of the zero temperature critical value
$V_{\rm c} \approx 1.32$. In both cases, charge order exists in an extended regime of filling near
$n=1/2$. Except at some special points (such as $V=1.4, n=1/2$), the order parameter of charge order
$ \left| n_{A} - n_{B} \right|$ changes non-monotonously with temperature. For $V=1.4$,
in the filling regime $0.39 < n < 0.5$ and near the reentrant transition temperature, we find PS
between a charge-disordered  and a charge-ordered phase.
Information on this PS, including the existence of a
metastable state and the first-order transition
line, is described by the "S"-shaped structure in the continuous $n-\mu$ curve.
At zero temperature, with increasing $n$, the charge-disordered ground state changes into a charge-ordered
one abruptly at $n=1/2$. Our analysis suggests that long-range Coulomb repulsion may destroy this PS
through its frustration effect. For $V =1.2$, the ground state is charge-disordered for all
fillings. The reentrant charge ordering transition is observed in the regime $ 0.42 < n <0.68$. It
becomes rather abrupt in the regime $ 0.5 <n < 0.65$.
Relevance of our results to the doped manganites is discussed.

\section{ Acknowledgement}

This work was supported by the Alexander von Humboldt foundation (N.-H. Tong),
the Research Grants Council of Hong Kong (Project No. HKU7088/01P) (S.-Q. Shen),
and by the DFG through SFB 484 (R. Bulla).


\begin{thebibliography}{99}

\bibitem{Cheong1} S. W. Cheong and H. Y. Hwang, in {\em Contribution to Colossal Magneto-resistance Oxides,
                  Monographs in Condensed Matter Science}. Edited by Y. Tokura (Gordon $\and$ Breach, London, 1999).
\bibitem{Kim1} K. H. Kim, S. Lee, T. W. Noh, and S. W. Cheong, cond-mat/0203150.
\bibitem{Tomioka1} Y. Tomioka, A. Asamitsu, H. Kuwahara, and Y. Tokura, J. Phys. Soc. Jpn {\bf 66},
                   302 (1997).
\bibitem{Kimura1} T. Kimura, R. Kumai, Y. Tokura, J. Q. Li, and Y. Matsui, Phys. Rev. B {\bf 58},
                  11081 (1998).
\bibitem{Li1} J. Q. Li, Y. Matsui, T. Kimura, and Y. Tokura, Phys. Rev. B {\bf 57}, R3205 (1998).
\bibitem{Ochiai1} A. Ochiai, T. Suzuki, and T. Kasuya, J. Phys. Soc. Jpn. {\bf 59}, 4129 (1990);
                  P. Fulde, B. Schmidt, and P. Thalmeier, Europhys. Lett. {\bf 31}, 323 (1995).
\bibitem{Vojta1} M. Vojta, A. H\"{u}bsch, R. M. Noack, Phys. Rev. B {\bf 63}, 045105 (2001).
\bibitem{Hubsch1} A. H\"{u}bsch, C. Waidacher, and K. W. Becker, Phys. Rev. B {\bf 64}, 241103 (2001).
\bibitem{Mori1} H. Mori, S. Tanaka, and T. Mori, Phys. Rev. B {\bf 57}, 12023 (1998).
\bibitem{Merino1} J. Merino and R. H. McKenzie, Phys. Rev. Lett. {\bf 87}, 237002 (2001).
\bibitem{Pietig1} R. Pietig, R. Bulla, and S. Blawid, Phys. Rev. Lett. {\bf 82}, 4046 (1999).
\bibitem{Kagan1} M. Y. Kagan, K. I. Kugel, and K. I. Khomskii, J. Exp. Theor. Phys. {\bf 93}, 415 (2001).
\bibitem{Tuan1} H. A. Tuan, Mod. Phys. Lett. B {\bf 15}, 1217 (2001).
\bibitem{Hellberg1} C. S. Hellberg, J. Appl. Phys. {\bf 89}, 6627 (2001).
\bibitem{Calandra1} M. Calandra, J. Merino, and R. H. McKenzie, Phys. Rev. B {\bf 66}, 195102 (2002).
\bibitem{Hoang1} A. T. Hoang and P. Thalmeier, J. Phys. Condens. Mat. {\bf 14}, 6639 (2002).
\bibitem{Sengupta1} P. Sengupta, A. W. Sandvik, and D. K. Campbell, Phys. Rev. B {\bf 65}, 155113 (2002);
                    M. Tsuchiizu and A. Furusaki, Phys. Rev. Lett. {\bf 88}, 056402 (2002);
		    M. Aichhorn, H. G. Evertz, W. von der Linden, and M. Potthoff, cond-mat/0402580. 
\bibitem{Ibarra1} M. R. Ibarra, G. M. Zhao, J. M. De Teresa, B. Garcia-Landa, Z. Arnold,
                  C. Marquina, P. A. Algarabel, H. Keller, and C. Ritter,
                  Phys. Rev. B {\bf 57}, 7446 (1998).
\bibitem{Liu1} H. L. Liu, S. L. Cooper, and S. W. Cheong, Phys. Rev. Lett. {\bf 81}, 4684 (1998).
\bibitem{Uhrig1} G. S. Uhrig and R. Vlaming, Phys. Rev. Lett. {\bf 71}, 271 (1993).
\bibitem{Caffarel1} M. Caffarel and W. Krauth, Phys. Rev. Lett. {\bf 72}, 1545 (1994).
\bibitem{Michielsen1} K. Michielsen and H. De Raedt, Phys. Rev. B {\bf 59}, 4565 (1999).
\bibitem{Dongen1} P. G. J. van Dongen, Phys. Rev. Lett. {\bf 74}, 182 (1995).
\bibitem{Zitzler1} R. Zitzler, Th. Pruschke, and R. Bulla, Eur. Phys. J. B {\bf 27}, 473 (2002).
\bibitem{Emery1} V. J. Emery, S. A. Kivelson, and H. Q. Lin,
Phys. Rev. Lett. {\bf 64}, 475 (1990).
\bibitem{White1} S. R. White and D. J. Scalapino, Phys. Rev. B {\bf 61}, 6320 (2000).
\bibitem{Freericks1}J. K. Freericks, E. H. Lieb, and D. Ueltschi, Phys. Rev. Lett. {\bf 88}, 106401 (2002);
                    J. K. Freericks, Ch. Gruber, and N. Macris, Phys. Rev. B {\bf 60}, 1617 (1999).
\bibitem{Shen1} S. Q. Shen and Z. D. Wang, Phys. Rev. B {\bf 58}, R8877 (1998); {\bf 59)}, 14484 (1999);
               {\bf 61}, 9532 (2000).
\bibitem{Yunoki1} S. Yunoki, A. Moreo, and E. Dagotto, Phys. Rev. Lett. {\bf 81}, 5612 (1998).
\bibitem{Moreo1} A. Moreo, S. Yunoki, and E. Dagotto, Science {\bf 283}, 2034 (1999).
\bibitem{Moreo2} A. Moreo, D. Scalapino, and E. Dagotto, Phys. Rev. B {\bf 43}, 11442 (1991).
\bibitem{Tong1} N. H. Tong and F. C. Pu, Phys. Rev. B {\bf 62}, 9425 (2000).
\bibitem{Tong2} N. H. Tong, S. Q. Shen, and F. C. Pu, Phys. Rev. B {\bf 64}, 235109 (2001).
\bibitem{Tong3} N. H. Tong and S. Q. Shen, Mod. Phys. Lett. B {\bf 15}, 1249 (2001).
\bibitem{Schiffer1} P. Schiffer, A. P. Ramirez, W. Bao, and S. W. Cheong, Phys. Rev. Lett. {\bf 75},
                   3336 (1995).
\bibitem{Babushkina1} N. A. Babushkina, L. M. Belova, D. I. Khomskii, K. I. Kugel, O. Yu. Gorbenko,
                      and A. R. Kaul, Phys. Rev. B {\bf 59}, 6994 (1999).
\bibitem{Anane1} A. Anane, J. P. Renard, L. Reversat, C. Dupas, P. Veillet, M. Viret, L. Pinsard,
                 and A. Revcolevschi, Phys. Rev. B {\bf 59}, 77 (1999).

\end{thebibliography}
\end{document}